\documentclass{article}
\usepackage{spconfa4,amsmath,graphicx,scalerel}
\usepackage{comment,multirow,multicol,balance,pifont,siunitx,circledsteps,scalerel,moresize,tipa,color,amsfonts,amssymb,setspace,bm,amsmath,xcolor}
\usepackage[
    hidelinks
]{hyperref}
\usepackage[export]{adjustbox}
\newcommand{\VECG}[1]{\boldsymbol{#1}} 
\title{Rethinking Language Model-Based Generative Speech Enhancement \\in the Latent Space of a Neural Audio Codec}
\name{Yihui Fu,
      Zhengyang Li,
      Tim Fingscheidt}
\address{Institute for Communications Technology, TU Braunschweig, Braunschweig, Germany}
\begin{document}
\ninept
\maketitle
\begin{abstract}
Language model (LM)-based speech enhancement (SE) has recently emerged rapidly using latent space features of neural audio codecs (NACs). In this paper, first, we present a unified framework covering \textit{six popular LM-based generative SE modeling paradigms} based on discrete/continuous latent NAC features: discrete or continuous autoregressive (\texttt{D/CAR}) SE, discrete or continuous non-autoregressive (\texttt{D/CNAR}) SE, discrete diffusion (\texttt{DDiff}) SE, and continuous flow matching (\texttt{CFM}) SE. Second, we are the first to compare their performance in a unified experimental setup and synopsis with diverse intrusive and non-intrusive metrics, enabling a fair and comprehensive evaluation. Third, we propose a fine-tuning strategy with \textit{auxiliary losses on reconstructed speech} to improve both intrusive and non-intrusive metrics. Trained and evaluated on URGENT 2025 Speech Enhancement Challenge data splits, all continuous-domain paradigms excel their discrete-domain counterparts. The overall best approach turns out to be \texttt{CNAR}. We further show that our proposed auxiliary loss fine-tuning strategy helps to improve DNSMOS, NISQA, PESQ, and POLQA consistently in all six paradigms\footnote{\url{https://github.com/felixfuyihui/AR_NAR_Diffusion_SE.git}}.

\end{abstract}
\section{Introduction}
\label{sec:intro}
Language model (LM)-based generative speech enhancement (SE) approaches have gained immense attention. Employing LMs in the enhancement process in a generative manner can harness their powerful semantic modeling capabilities, enabling better speech reconstruction beyond simple noise removal. LM-based SE typically takes discrete tokens or continuous latent features of the noisy speech as input to predict the corresponding clean speech features. Autoregressive (AR), non-autoregressive (NAR), diffusion (typically NAR), and flow matching (typically NAR) are the most popular modeling paradigms in terms of LM-based generative SE. 

Concretely, AR SE takes noisy speech tokens and optionally noisy speech semantic features as the prompt to predict clean speech tokens in an AR manner. An embedding layer transforms the discrete tokens into continuous embeddings, followed by causal transformer-based modules (e.g., \texttt{Llama} \cite{touvron2023llama}), is commonly adopted \cite{yao2025gense, kang2025llase,yan2025unise,kammoun2025modeling,lanzendorfer2025high}. NAR SE takes the noisy speech tokens or continuous latent features as the input to generate the entire sequence of the corresponding clean speech features at once. A non-causal transformer-based architecture is commonly adopted \cite{li2025speech}. Diffusion SE aims at modeling two processes between the clean speech distribution and the noisy signal prior distribution. The forward process transforms the clean features distribution into a known noisy features prior distribution, while the reverse process aims at sampling the clean features distribution starting from the prior distribution \cite{zhang2025anyenhance,yang2024genhancer,fu2026discontse}. Flow matching SE \cite{lipmanflow, welker2025flowdec} aims at estimating a vector field to transform a prior distribution (noisy speech features) into a target data distribution (clean speech features). Solving an ordinary differential equation (ODE) during inference is adopted to sample the enhanced speech feature.

However, \textit{the research community lacks a broad and fair comparison of all these LM-based generative SE paradigms under a unified experimental setup and metrics}. Furthermore, while achieving promising results on non-intrusive metrics, their performance on intrusive metrics is often neither reported nor discussed.

On the other hand, neural audio codecs (NACs) are being widely adopted in LM-based SE, benefiting from their compressed representations of noisy and clean speech for sequence modeling. Recent successful and widely adopted NACs are \texttt{EnCodec} \cite{defossezhigh} (at 1.5, 3, 6, and 12 kbps at 24 kHz waveform), \texttt{DAC} \cite{kumar2023high} (at 6 kbps at 16 kHz waveform), \texttt{BiCodec} \cite{wang2025spark} (at 0.65 kbps at 16 kHz waveform), and \texttt{X-Codec} \cite{ye2025codec} (at 0.5, 1, 1.5, 2, and 4 kbps at 16 kHz waveform), etc. Features in the latent space, including continuous features from NACs' encoders and discrete tokens from NACs' residual vector quantization (RVQ)-based quantizer \cite{rvq}, are commonly adopted in LM-based generative SE.

Our contributions in this paper are threefold. First, we propose a unified decoder-only LM framework for the six most popular LM-based generative SE modeling paradigms based on latent NAC features, namely discrete/continuous autoregressive (\texttt{D/CAR}) SE, discrete/continuous non-autoregressive (\texttt{D/CNAR}) SE, discrete diffusion (\texttt{DDiff}) SE in NAR manner, and continuous flow matching (\texttt{CFM}) SE in NAR manner. Second, we deliver a comprehensive and fair evaluation of the paradigms above, employing \textit{both} non-intrusive and intrusive metrics, which has rarely been done in LM-based generative SE research so far. \textit{Our research claims to be the first synopsis of (non-) intrusive metrics of a range of LM-based SE paradigms}. Third, a fine-tuning strategy with auxiliary losses on reconstructed speech is proposed to further improve their performance, consistently showing better DNSMOS, NISQA, PESQ, and POLQA. Note that the fine-tuning only applies to the LM model parameters, requiring no updates to the pretrained codec model. This ensures the fine-tuned model to retain a high versatility.

\begin{figure*}[ht!]
    \centering
    \raisebox{0cm}{\hspace{-0.32cm}}
    \includegraphics[width=2.00\columnwidth]{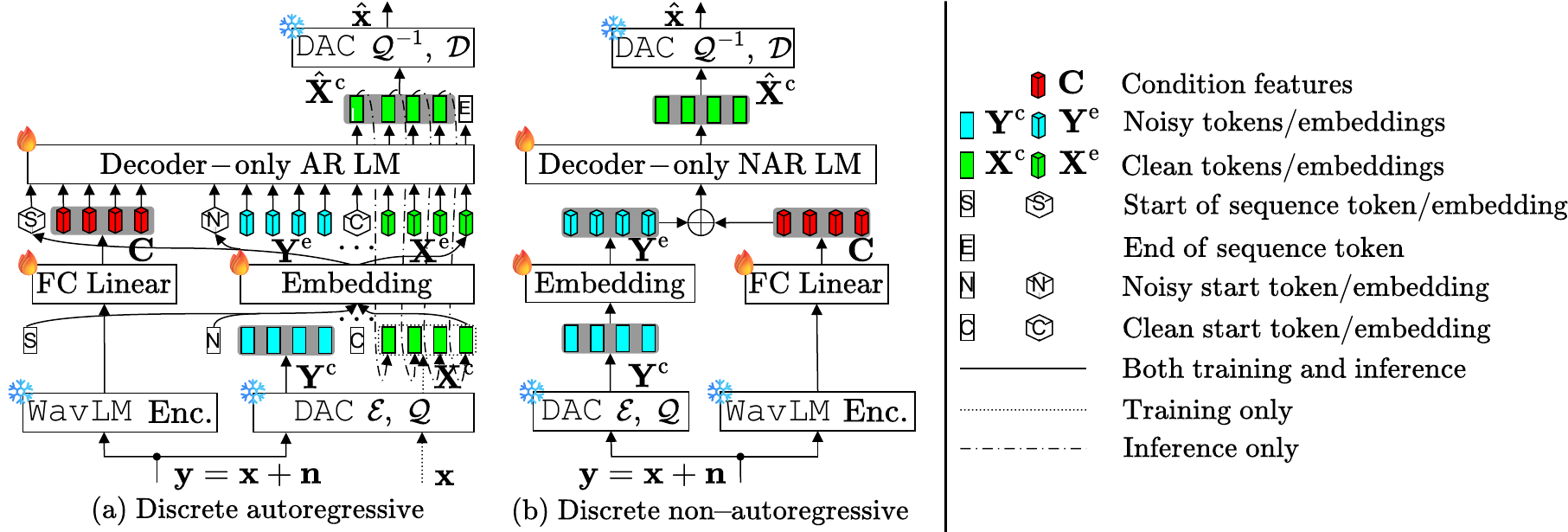}
    \caption{Architecture
    of the \textbf{proposed} (a) discrete autoregressive (\texttt{DAR}) SE and (b) discrete non-autoregressive (\texttt{DNAR}) SE. Continuous (non-) autoregressive (\texttt{C(N)AR}) SE, discrete diffusion (\texttt{DDiff}) SE, and continuous flow matching (\texttt{CFM}) share similar topologies with the corresponding discrete domain SE, with slight modifications as described in the text.
    }
    \label{fig:ar_nar}
\end{figure*}

\section{Methods}
\label{sec:format}

\subsection{Neural Audio Codecs}
Neural audio codecs (NACs) \cite{defossezhigh, kumar2023high, wang2025spark, ye2025codec} learn to map a time-domain signal into compact sequences of discrete tokens via a residual vector quantization (RVQ) scheme \cite{rvq}. Letting $\mathbf{y}$ and $\mathbf{x}$ represent the time-domain noisy and clean speech vectors, respectively, the tokenization and detokenization process of a NAC is:

\begin{equation}
\label{tokenize}
\mathbf{x}\xrightarrow{\mathcal{E}}{\mathbf{X}}\in\mathbb{R}^{D\times L}\xrightarrow{\mathcal{Q}}\mathbf{X}^{\mathrm{c}}=(X^{\mathrm{c}}_\ell)\in\mathbb{N}^{C\times L}=\{0,...,N\!-\!1\}^{C\times L},
\end{equation}
\begin{equation}
\label{detokenize}
\mathbf{X}^{\mathrm{c}}\xrightarrow{\mathcal{Q}^{-1}}\tilde{\mathbf{X}}\in\mathbb{R}^{D\times L}\xrightarrow{\mathcal{D}}\tilde{\mathbf{x}}, 
\end{equation}
where $\mathcal{E}$ and $\mathcal{D}$ represent the NAC encoder and decoder, respectively, $\mathcal{Q}$ and $\mathcal{Q}^{-1}$ represent the quantization and dequantization processes, respectively, $\mathbf{X}$ and $\mathbf{X}^{\mathrm{c}}$ represent the continuous features from encoder and discrete tokens from the quantizer, with sequence length $L$ in frames, encoder output/decoder input feature dimension $D$, number of codebooks $C$, and codebook size $N$, respectively. The same process can be applied to $\mathbf{y}$. In this paper, we select the popular \texttt{DAC} \cite{kumar2023high} as the NAC model, with 20 ms non-overlapping frames on a 16 kHz waveform, which is further justified in Section \ref{sec:ablation}.
\vspace{-6pt}
\subsection{Condition Features Extractor}
We adopt the pre-trained \texttt{WavLM} \cite{chen2022wavlm} encoder as the condition feature extractor to provide auxiliary semantic information. \texttt{WavLM} is a self-supervised speech representation model designed to learn robust and generalizable audio features from large-scale unlabeled speech data. Benefiting from its 20 ms non-overlapping frames on a 16 kHz waveform, the condition features extracted by the \texttt{WavLM} encoder are time-aligned with \texttt{DAC} features. 

\vspace{-6pt}
\subsection{LM-Based Generative SE Modeling Paradigms}
\textbf{(i) Discrete Autoregressive (\texttt{DAR}) SE} \\
As shown in Fig.~\ref{fig:ar_nar} (a), during training, the \texttt{WavLM} encoder and a fully connected (FC) linear layer extract the frame-by-frame condition features $\mathbf{C} \in\mathbb{R}^{H\times L}$ from noisy speech $\mathbf{y}$, where $H$ is the hidden dimension of the subsequent LM. \texttt{DAC} tokens of both noisy and clean speech $\mathbf{Y}^\mathrm{c},\mathbf{X}^\mathrm{c}\in\mathbb{N}^{C\times L}$ are extracted as the input to the $C$ parallel embedding layers with embedding dimension $H$. The sum of the $C$ output embeddings are $\mathbf{Y}^{\mathrm{e}},\mathbf{X}^{\mathrm{e}}\in\mathbb{R}^{H \times L}$, which are appended to $\mathbf{C}$ along the sequence length dimension as the input to the decoder-only AR LM to estimate the $C$ logits/tokens for each frame. Unlike the token flattening strategy in \cite{lanzendorfer2025high}, which increases the sequence length by a factor of $C$ times, our strategy ensures that the sequence length remains unchanged to improve training and inference efficiency. Note that a set of special tokens/embeddings is employed to segment different domain features, including start of sequence token/embedding (\includegraphics[height=1em, valign=c, raise=-0.1ex]{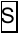}/\includegraphics[height=1em, valign=c, raise=-0.1ex]{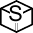}), end of sequence token (\includegraphics[height=1em, valign=c, raise=-0.1ex]{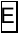}), noisy start token/embedding (\includegraphics[height=1em, valign=c, raise=-0.1ex]{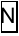}/\includegraphics[height=1em, valign=c, raise=-0.1ex]{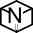}), clean start token/embedding (\includegraphics[height=1em, valign=c, raise=-0.1ex]{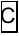}/\includegraphics[height=1em, valign=c, raise=-0.1ex]{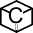}). The LM is trained in a teacher-forcing manner, and the cross-entropy (CE) loss between the predicted logits and the clean tokens $\mathbf{X}^{\mathrm{c}}$ is minimized. The LM, together with FC linear and embedding layers, is updated, while the \texttt{WavLM} encoder and \texttt{DAC} modules are frozen. During inference, ${\mathbf{C}}$ and $\mathbf{Y}^{\mathrm{e}}$ are concatenated as prompt to the LM to predict the enhanced tokens $\hat{\mathbf{X}}^{\mathrm{c}}$ in a step-by-step manner, which are further input to the \texttt{DAC} dequantizer $\mathcal{Q}^{-1}$ and decoder $\mathcal{D}$ to estimate the enhanced speech waveform vector $\hat{\mathbf{x}}$. \\
\textbf{(ii) Continuous Autoregressive (\texttt{CAR}) SE} \\
During training, the \texttt{DAC} dequantizer output of noisy and clean speech $\tilde{\mathbf{Y}}$ and $\tilde{\mathbf{X}}$ are extracted, followed by an FC linear layer (substitution of the embedding layers in \texttt{DAR} in Fig.~\ref{fig:ar_nar} (a) to estimate embeddings $\mathbf{Y}^{\mathrm{e}},~\mathbf{X}^{\mathrm{e}}\in\mathbb{R}^{H \times L}$, which are appended to \texttt{WavLM} condition features $\mathbf{C}$ along the sequence length dimension as the input to the decoder-only AR LM. The LM estimates the enhanced \textit{unquantized} features $\hat{\mathbf{X}}$. Minimizing the mean absolute error (MAE) loss between $\hat{\mathbf{X}}$ and clean speech \texttt{DAC} encoder output $\mathbf{X}$ is adopted during training in a teacher-forcing manner, which corresponds to minimizing the CE loss in the discrete token domain \cite{kammoun2025modeling}. The model updating/freezing strategies are the same as with \texttt{DAR}. During inference, ${\mathbf{C}}$ and $\tilde{\mathbf{Y}}$ are concatenated as the prompt to estimate $\hat{\mathbf{X}}$ in a step-by-step manner. It is noticed that in both training and inference, we always use the \textit{dequantized} representations $\tilde{\mathbf{Y}}$ and $\tilde{\mathbf{X}}$ as the input to the LM, instead of the \textit{unquantized} \texttt{DAC} encoder output features $\mathbf{Y}$ and $\mathbf{X}$. This can mitigate the error accumulation problem in the AR model \cite{kammoun2025modeling}. For the same reason, in inference, the LM predicts the unquantized feature $\hat{\mathbf{X}}$, which is further quantized ($\mathcal{Q}$) and dequantized ($\mathcal{Q}^{-1}$) to $\hat{\tilde{\mathbf{X}}}$ as input for the next autoregressive step. The estimated $\hat{\tilde{\mathbf{X}}}$ is further input to the \texttt{DAC} decoder $\mathcal{D}$ to estimate the enhanced speech waveform vector $\hat{\mathbf{x}}$. \\
\textbf{(iii) Discrete Non-Autoregressive (\texttt{DNAR}) SE} \\
As shown in Fig.~\ref{fig:ar_nar} (b), \texttt{DNAR} aims to perform SE in the discrete token domain in a NAR manner. 
During training, the condition features $\mathbf{C}$ and the noisy speech embeddings $\mathbf{Y}^{\mathrm{e}}$ are extracted and added as the input to the decoder-only NAR LM. The CE loss between the predicted logits and the clean tokens $\mathbf{X}^{\mathrm{c}}$ is minimized. The module updating/freezing strategies are the same as with \texttt{DAR}. During inference, the same pipeline is adopted to estimate all the enhanced tokens $\hat{\mathbf{X}}^{\mathrm{c}}$ simultaneously in a single forward process, which are further input to the \texttt{DAC} dequantizer $\mathcal{Q}^{-1}$ and decoder $\mathcal{D}$ to estimate the enhanced speech waveform vector $\hat{\mathbf{x}}$. \\
\textbf{(iv) Continuous Non-Autoregressive (\texttt{CNAR}) SE} \\
During training, an FC linear layer (substitution of the embedding layers in \texttt{DNAR} in Fig. \ref{fig:ar_nar} (b) is applied to \texttt{DAC} encoder output $\mathbf{Y}$ to estimate $\mathbf{Y}^{\mathrm{e}}$, which is appended to $\mathbf{C}$ as input to the decoder-only NAR LM. The MAE loss between the enhanced features $\hat{\mathbf{X}}$ and the clean \texttt{DAC} encoder output $\mathbf{X}$ is minimized. The module updating/freezing strategies are the same as with \texttt{DNAR}. During inference, the same pipeline is adopted to estimate all the enhanced features $\hat{\mathbf{X}}$ simultaneously in a single forward process, which are further input to the \texttt{DAC} quantizer $\mathcal{Q}$, dequantizer $\mathcal{Q}^{-1}$, and decoder $\mathcal{D}$ to estimate the enhanced speech waveform vector $\hat{\mathbf{x}}$. \\
\textbf{(v) Discrete Diffusion (\texttt{DDiff}) SE} \\
\texttt{DDiff} largely follows the pattern of \texttt{DNAR} in Fig.~\ref{fig:ar_nar} (b). During training, \texttt{MaskGIT} \cite{chang2022maskgit} is adopted as a random mask generator to apply masking on the clean tokens $\mathbf{X}^{\mathrm{c}}$ with a special mask token $[\mathrm{M}]$ with index value $N$ and time step index $t \backsim \mathcal{U}(0,1)$ to generate the masked token sequence $\mathbf{X}^{\mathrm{c}}_t$. Then, the embeddings of $\mathbf{X}^{\mathrm{c}}_t$, namely $\mathbf{X}_t^\mathrm{e}$, are appended to the condition features ${\mathbf{C}}$ as the input to the decoder-only NAR LM. The CE loss between the predicted logits and the clean tokens $\mathbf{X}^{\mathrm{c}}$ \textit{on the masked positions} is minimized. The model training strategy is the same as for \texttt{DNAR}. During inference, a fully masked sequence of tokens $\mathbf{X}^{\mathrm{c}}_T\in\mathbb{N}^{C\times L}$ with time step index $T\!=\!1.0$ is taken as the initial state of input to the model to predict the tokens on the masked positions. Then, an $N'$-iteration diffusion process is applied. For each iteration, a portion of entry positions of the model output tokens, which are considered as estimated low-confidence tokens \cite{chang2022maskgit}, are remasked by $[\mathrm{M}]$ as the next step's input semantic token sequence $\mathbf{X}^{\mathrm{c}}_{t-\Delta t}$, with $\Delta t=T/N'$ and $t\in\{T,T\!-\!\Delta t\,...,\Delta t\}$. More details about inference are in \cite{fu2026discontse}. The last-step output $\mathbf{X}^{\mathrm{c}}_{0}$ is input to the \texttt{DAC} dequantizer $\mathcal{Q}^{-1}$ and decoder $\mathcal{D}$ to estimate the enhanced speech waveform vector $\hat{\mathbf{x}}$. \\
\textbf{(vi) Continuous Flow Matching (\texttt{CFM}) SE} \\
\texttt{CFM} \cite{lipmanflow,welker2025flowdec} largely follows the pattern of \texttt{CNAR} in Fig.~\ref{fig:ar_nar} (b) and is applied in the \texttt{DAC} encoder output feature domain. During training, the state of the encoder output $\mathbf{X}$ with time step index $t\backsim \mathcal{U}(0,1)$ is calculated by
\begin{equation}
\mathbf{X}_{t}=t\cdot \mathbf{X}_{1} + (1-t)\cdot \mathbf{X}_{0},
\end{equation}
where $\mathbf{X}_{1}=\mathbf{X}$, $\mathbf{X}_{0}=\mathbf{Y}$. An FC linear layer (substitution of the embedding layers in \texttt{DNAR} in Fig.~\ref{fig:ar_nar} (b) is applied to $\mathbf{X}_{t}$ to output $\mathbf{X}_{t}^{\mathrm{e}}$, which is appended to $\mathbf{C}$ as the input to the decoder-only NAR LM to estimate the velocity $\mathbf{v}_{ \VECG\theta}(\mathbf{X}_{1},\mathbf{X}_{0}, \mathbf{C},t)$, where $\VECG\theta$ are the trainable parameters of \texttt{CFM}. The FM loss 
\begin{equation}
J^{\textrm{fm}} = ||\mathbf{v}_{\VECG\theta}(\mathbf{X}_{1},\mathbf{X}_{0}, \mathbf{Y}^\mathrm{c},t)-(\mathbf{X}_{1}-\mathbf{X}_{0})||,
\end{equation}
is adopted. The model updating/freezing strategies are the same as with \texttt{CNAR}. During inference, starting from $\!T=\!0.0$, an ordinary differential equation (ODE) with $N'$ iterations is adopted:
\begin{equation}
\mathbf{X}_{t+\Delta t}=\mathbf{X}_{t}+\mathbf{v}_{\VECG\theta}(\mathbf{X}_{1},\mathbf{X}_{0}, \mathbf{Y}^\mathrm{c},t)\cdot\Delta t,
\end{equation}
where $\Delta t\!=\!1.0/N'$, $t\in\{0.0,\Delta t,...,1.0\!-\!\Delta t\}$. Finally, the estimated $\mathbf{X}_{1}$ is input to the \texttt{DAC} quantizer $\mathcal{Q}$, dequantizer $\mathcal{Q}^{-1}$, and decoder $\mathcal{D}$ to estimate the enhanced speech waveform vector $\hat{\mathbf{x}}$.

\vspace{-6pt}
\subsection{Fine-tuning with Auxiliary Losses}
To improve the performance of generative SE models, after the pre-training of the six paradigms above, we propose a fine-tuning strategy with auxiliary losses on reconstructed speech:
\begin{equation}
\label{ft}
J=J^{\mathrm{task}} + J^{\mathrm{aux}}(\hat{\mathbf{x}}, \mathbf{x}),
\end{equation}
\begin{equation}
\label{aux}
\begin{aligned}
J^{\mathrm{aux}}(\hat{\mathbf{x}}, \mathbf{x}) &= \alpha \cdot J^{\mathrm{Braun}}(\mathrm{STFT}(\hat{\mathbf{x}}), \mathrm{STFT}(\mathbf{x})) +\gamma \cdot J^{\mathrm{L1}}(\hat{\mathbf{x}}, \mathbf{x}) \\
&+ \delta  \cdot J^{\mathrm{PESQ}}(\hat{\mathbf{x}}, \mathbf{x}) + \eta  \cdot J^{\mathrm{STOI}}(\hat{\mathbf{x}}, \mathbf{x}),
\end{aligned}
\end{equation}
where $J^{\mathrm{task}}$ is the corresponding CE or MAE loss for each paradigm, $J^{\mathrm{Braun}}$ is Braun's and Tashev's loss in the frequency domain following \cite{braun2021consolidated,effcrn}, $J^{\mathrm{L1}}$ is the time-domain L1 loss, $J^{\mathrm{PESQ}}$ and $J^{\mathrm{STOI}}$ are differentiable PESQ \cite{torch-pesq} and STOI \cite{pytorch_stoi} losses, respectively, $\mathrm{STFT}(\cdot)$ represents the short-term Fourier transform with a frame length of 512 samples and 25\% frameshift, $\alpha$, $\gamma$, $\delta$, $\eta$ are hyperparameters. The module updating/freezing strategies are the same as the corresponding pre-training strategies. It is noticed that we use a straight-through estimator (STE) \cite{yin2018understanding} in discrete-domain paradigms when estimating tokens from logits to enable gradient backpropagation.
\vspace{-6pt}
\section{Experimental Setup}
\vspace{-3pt}
\subsection{Networks}
In the proposed SE models, the pretrained 16 kHz \texttt{DAC} \cite{descript-audio-codec} and \texttt{WavLM} \cite{wavlm-large} weights are adopted. For \texttt{DAC}, the encoder output dimension is $D\!=\!1024$, the codebook size is $N\!=\!1024$, and the number of codebooks is $C\!=\!12$, respectively. For \texttt{WavLM}, the 6th encoder layer's output is adopted \cite{yao2025gense} with dimension $D\!=\!1024$.
For the AR and NAR tasks, we use a 24-layer \texttt{Llama} \cite{touvron2023llama} and non-causal transformer \cite{vaswani2017attention}, respectively, with both hidden dimension $H\!=\!1024$ and 16 heads. For fair comparison, all \texttt{Llama} and transformer models are trained from scratch, thereby not using any pre-trained models. The total number of trainable parameters in the proposed \texttt{DAR}, \texttt{CAR}, \texttt{DNAR}, \texttt{CNAR}, \texttt{DDiff}, and \texttt{CFM} are 278.0 M, 256.0 M, 227.8 M, 204.7 M, 227.8 M, and 204.7 M, respectively, while the number of frozen parameters of the \texttt{DAC} and \texttt{WavLM} encoders is 74.2 M and 158.3 M, respectively.

\vspace{-8pt}
\subsection{Data and Metrics}

\label{results}
\begin{table}[t!]
\caption{Ablation study on popular NACs (on clean speech) and SE inference strategies of \texttt{DAR}, \texttt{DDiff}, and \texttt{CFM} (on noisy speech) on $\mathcal{D}^{\textrm{val}}$. For \texttt{DAR}, greedy search (GS) and beam search (BS) with different beam sizes ($B$) are explored. For \texttt{DDiff} and \texttt{CFM}, different numbers of iterations ($N'$) are explored. Selected setups in italic.}
{
\scalebox{0.93}
{
\begin{tabular}{@{\hskip 3pt}l@{\hskip 3pt}|@{\hskip 3pt}c@{\hskip 3pt}|@{\hskip 3pt}c@{\hskip 3pt}c@{\hskip 3pt}c@{\hskip 3pt}c@{\hskip 3pt}c@{\hskip 3pt}}
\hline
Method    & Strategy & \begin{tabular}[c]{@{}c@{}}DNS\\ MOS\end{tabular} & NISQA & UTMOS & PESQ & ESTOI   \\\hline\hline 
Clean     &-   &3.05    &3.64  &2.81  &-  &-          \\
~-\texttt{DAC} \cite{kumar2023high}     &-   &\textit{3.08}    &\textit{3.68}  &\textit{2.79}  &\textit{3.83}  &\textit{0.93}         \\
~-\texttt{BiCodec} \cite{wang2025spark}  &- &3.21 &3.74 &3.11  &2.18  &0.77  \\
~-\texttt{X}-\texttt{Codec}-1 \cite{ye2025codec}  &- &3.27 &3.77 &3.17  &1.64  &0.65  \\
~-\texttt{X}-\texttt{Codec}-8 \cite{ye2025codec}  &- &3.19 &4.04 &3.11  &2.61  &0.78  \\\hline\hline
Noisy     &- &1.78    &1.47  &1.51  &1.26  &0.57   \\\hline\hline
\multirow{3}{*}{\texttt{DAR}}     & GS   &\textit{2.74}    &\textit{2.60}  &\textit{2.03}  &\textit{1.50}  &\textit{0.56}      \\
     & BS, $B=3$   &2.74    &2.60  &2.03  &1.50  &0.56 \\
     & BS, $B=5$   &2.74    &2.61  &2.04  &1.50  &0.56 \\\hline\hline
\multirow{4}{*}{\texttt{DDiff}}     & \hspace{-5pt}$N'=1$   &2.69    &2.87  & 2.04  &1.54  &0.63   \\
    & \hspace{-5pt}$N'=3$   &\textit{2.79}    &\textit{2.97}  &\textit{2.14}  &\textit{1.59}  &\textit{0.64}         \\
     & $N'=10$   &2.79    &2.98  &2.15  &1.59  &0.63          \\
     & $N'=20$   &2.77    &2.94  &2.16  &1.57  &0.63          \\ \hline\hline
\multirow{4}{*}{\texttt{CFM}}     & \hspace{-5pt}$N'=1$   &2.96    &3.22  &2.34  &1.89  &0.69          \\
     & \hspace{-5pt}$N'=3$   &\textit{2.99}    &\textit{3.63}  &\textit{2.35}  &\textit{1.89}  &\textit{0.69}          \\
     & $N'=10$   &2.98    &3.64  &2.27  &1.84  &0.69          \\
     & $N'=20$   &2.96    &3.61  &2.25  &1.82  &0.68          \\ \hline

\end{tabular}
}
}
\label{tab:ablation}
\end{table}

We train and evaluate our proposed LM-based SE models on the large-scale diverse-source URGENT 2025 Speech Enhancement Challenge data splits \cite{saijo25_interspeech}, however, excluding CommonVoice 19.0 \cite{CommonVoice-Ardila2020} due to its occasional background noise. For training data $\mathcal{D}^{\textrm{train}}$, we mainly follow the strategies in the challenge \cite{saijo25_interspeech}, in which six kinds of distortions are adopted: additive noise, reverberation, clipping, codec loss (MP3 and OGG), packet loss, and wind noise. We do not adopt the bandwidth limitation distortion from the challenge. The signal-to-noise ratio (SNR) range is [-5, 20] dB. An active speech level with a range [-36, -16] dB is adopted before mixing speech and noise data \cite{ITUSRN}. All waveforms are downsampled to 16 kHz. We configure a 1202.2 h training set $\mathcal{D}^{\textrm{train}}$. We adopt the officially published validation and non-blind test set of the challenge and downsample it to 16 kHz, leading to our validation set $\mathcal{D}^{\textrm{val}}$ and test set $\mathcal{D}^{\textrm{test}}$. Each contains 1000 waveforms.

We employ intrusive SE metrics, including PESQ \cite{rix2001perceptual}, and ESTOI \cite{jensen2016algorithm}, non-intrusive SE metrics, including DNSMOS \cite{reddy2022dnsmos}, NISQA \cite{mittag2021nisqa}, and UTMOS \cite{saeki22c_interspeech}, downstream-task-independent metrics, including Levenshtein phone similarity (LPS ~=~1~-~Levenshtein phone distance (LPD)) \cite{pirklbauer2023evaluation}, as a language-independent phone fidelity metric being sensitive to hallucinations.

\subsection{Model Training}

The proposed LM-based SE models are trained on 4 \texttt{Nvidia H100}, using a batch size of 8 with a 5-second waveform in each mini-batch, for 150 K steps, taking approximately 2 days. An AdamW optimizer with $\beta=(0.9, 0.95)$ and weight decay 0.05 is adopted. The learning rate is 0.0001 with 5 K steps of warmup. For fine-tuning (-FT) with auxiliary losses, the models are trained for 50 K further steps, with $\alpha\!=\!10.0$, $\gamma\!=\!20.0$, $\delta\!=\!0.5$, and $\eta\!=\!1.0$. The learning rate is 0.00001 with 5 K steps of warmup. For a fair comparison on $\mathcal{D}^{\textrm{test}}$, the non-fine-tuned models are also further trained for 50 K steps.

\begin{table}[t!]
\centering
\caption{Performance comparison of LM-based SE methods using \texttt{DAC} on $\mathcal{D}^{\textrm{test}}$, and corresponding fine-tuning (-FT) including an auxiliary loss (\ref{ft}). The best performance in each of the last two table segments for each metric is in bold; the second-best is underlined.}
\scalebox{0.94}
{\hspace{-0.18cm}
{
\begin{tabular}{@{\hskip 3pt}l|c@{\hskip 3pt}c@{\hskip 3pt}c@{\hskip 3pt}c@{\hskip 3pt}c@{\hskip 3pt}c@{\hskip 3pt}c@{\hskip 3pt}}
\hline
Method     & \begin{tabular}[c]{@{}c@{}}DNS\\ MOS\end{tabular} & NISQA & UTMOS & PESQ  & POLQA & ESTOI  & LPS  \\ \hline\hline
Noisy   &1.84 &1.65 &1.56 &1.31 &1.85 &0.61 &0.62  \\
Clean   &2.93 &3.33 &2.50 &-  &- &- &-  \\ 
~-\texttt{DAC} \cite{kumar2023high}   &2.99  &3.39  &2.45  &3.85  &4.24 &0.93   &0.93   \\
\hline\hline
\texttt{DAR}    &2.71  &2.63  &1.96  &1.50 &1.99  &0.58  &0.66   \\
\texttt{CAR}    &\underline{2.95}  &3.20  &\underline{2.27}  &\underline{1.83}    &\underline{2.61} &0.66  &\underline{0.73}   \\
\texttt{DNAR}   &2.67  &2.82  &1.97  &1.55  &2.11 &0.66  &0.69   \\ 
\texttt{CNAR}   &\textbf{3.01}  &\underline{3.39}  &\textbf{2.40}  &\textbf{1.99} &\textbf{2.75} &\textbf{0.75}  &\textbf{0.76}   \\ 
\texttt{DDiff}   &2.75  &2.93  &2.05  &1.58  &2.17 &0.66  &0.68     \\
\texttt{CFM}   &\underline{2.95}  &\textbf{3.46}  &2.15  &1.81  &2.50 &\underline{0.71}  &0.70     \\\hline\hline
\texttt{DAR}-FT   &2.73  &2.73  &1.95  &1.61 &2.09 &0.56  &0.63   \\
\texttt{CAR}-FT   &\underline{3.02}  &3.33  &\underline{2.28}  &1.99   &2.66  &0.64  &0.71   \\
\texttt{DNAR}-FT  &2.72  &2.88  &1.98  &1.84 &2.32 &0.66  &0.67      \\ 
\texttt{CNAR}-FT   &\textbf{3.03}  &\underline{3.41}  &\textbf{2.38}  &\textbf{2.41} &\textbf{3.00} &\textbf{0.76}  &\textbf{0.76}     \\ 
\texttt{DDiff}-FT   &2.80  &3.01  &2.08  &1.81  &2.36 &0.67  &0.68    \\
\texttt{CFM}-FT   &2.96  &\textbf{3.53}  &2.25  &\underline{2.15}  & \underline{2.76} &\underline{0.73}  &\underline{0.72}    \\\hline
\end{tabular}
}
}
\label{tab:performance}
\end{table}

\vspace{-8pt}
\section{Results and Discussion}
\vspace{-4pt}
\subsection{Ablation on the Validation Set} \label{sec:ablation}
In Table \ref{tab:ablation}, we perform an ablation study on $\mathcal{D}^{\mathrm{val}}$. We first apply the recent popular NACs to the clean speech, including \texttt{DAC} \cite{kumar2023high} with 12 codebooks, the single-codebook \texttt{BiCodec} \cite{wang2025spark}, \texttt{X}-\texttt{Codec} with 1 and 8 codebooks (\texttt{X}-\texttt{Codec}-1 and \texttt{X}-\texttt{Codec}-8), respectively.

Concerning the non-intrusive metrics (DNSMOS, NISQA, and UTMOS), we observe that all NACs' performance is in the range or even better than the uncoded (clean) waveform. Concerning the intrusive metrics PESQ and ESTOI, we observe \texttt{X}-\texttt{Codec}-1 to be the poorest, followed by \texttt{BiCodec} and then the \texttt{X}-\texttt{Codec}-8. With a significant margin, \texttt{DAC} is the clear winner of this ablation. So we choose it for all further experiments in Tables \ref{tab:ablation} and \ref{tab:performance}.

In the lower three segments of Table \ref{tab:ablation}, we investigate different hyperparameter setups during the inference of \texttt{DAR}, \texttt{DDiff}, and \texttt{CFM}.
While \texttt{DAR} is anyway quite robust to hyperparameter choices, we observe for \texttt{DDiff} and \texttt{CFM} that $N'=3$ iterations already deliver the best results. Chosen settings are in italic.

\subsection{Performance Analysis on the Test Set}
Table \ref{tab:performance} shows the performance comparison among the proposed six paradigms (second table segment) and the corresponding fine-tuning (-FT) results (third table segment). 

\textit{Across all metrics, we observe that the continuous-domain paradigms (methods named ``\texttt{C}...") are consistently outperforming their discrete-domain counterparts (methods named ``\texttt{D}...")}. 
This is due to predicting $D=1024$ continuous features can offer greater fault tolerance than predicting $C=12$ discrete tokens per frame, and quantization improves the likelihood of correct token recovery.
Among the discrete approaches, \texttt{DAR} is the worst and \texttt{DDiff} is clearly the best. All of them are excelled by \texttt{CFM}, then \texttt{CAR}, and the final winning approach \texttt{CNAR} is top in all but one metric and second in the other. \textit{It is impressive to see that \texttt{CNAR} is leading with a significant margin vs. all the other methods in quite some metrics}, 
benefiting from its totally matched training/inference pattern.
 
In the bottom segment of Table \ref{tab:performance}, we observe the very same rank orders, with continuous-domain paradigms in the lead, with method \texttt{CNAR}-FT being clearly top-ranked. \textit{Interestingly, the fine-tuning is able to consistently improve non-intrusive metrics (DNSMOS and NISQA) and intrusive metrics (PESQ and POLQA).} 

\vspace{-6pt}
\section{Conclusions}
\vspace{-4pt}
In this paper, we present a unified framework for six popular decoder-only LM-based generative SE modeling paradigms based on discrete/continuous latent NAC features. We are the first to deliver a comprehensive and fair synopsis of the paradigms by employing both non-intrusive and intrusive metrics. Continuous-domain paradigms excel discrete ones, with a continuous non-autoregressive approach being clearly on top. Our proposed fine-tuning strategy consistently improves DNSMOS, NISQA, PESQ, and POLQA across all six paradigms.

\vspace{-4pt}
\section{Acknowledgment}
Computational resources were provided by the German AI Service Center WestAI.


\clearpage
\begin{spacing}{0.9} 
\bibliographystyle{IEEEbib}
\bibliography{refsetal}
\end{spacing}   
\end{document}